# Anisotropic Thermal Conductivity of 4H and 6H Silicon Carbide Measured Using Time-Domain Thermoreflectance

Xin Qian, Puqing Jiang, and Ronggui Yang[*]

Department of Mechanical Engineering

University of Colorado, Boulder, CO 80309, USA

**Abstract**

Silicon carbide (SiC) is a wide bandgap (WBG) semiconductor with promising applications in high-power and high-frequency electronics. Among its many useful properties, the high thermal conductivity is crucial. In this letter, the anisotropic thermal conductivity of three SiC samples, n-type 4H-SiC (N-doped $1\times10^{19}$ cm$^{-3}$), unintentionally doped (UID) semi-insulating (SI) 4H-SiC, and SI 6H-SiC (V-doped $1\times10^{17}$ cm$^{-3}$), is measured using femtosecond laser based time-domain thermoreflectance (TDTR) over a temperature range from 250 K to 450 K. We simultaneously measure the thermal conductivity parallel to ($k_r$) and across the hexagonal plane ($k_z$) for SiC by choosing the appropriate laser spot radius and the modulation frequency for the TDTR measurements. For both $k_r$ and $k_z$, the following decreasing order of thermal conductivity value is observed: SI 4H-SiC > n-type 4H-SiC > SI 6H-SiC. This work serves as an important benchmark for understanding thermal transport in WBG semiconductors.

Excellent properties of silicon carbide (SiC) including its high electron mobility,[1-2] wide electronic bandgap,[3-4] and superior chemical stability [5] have led to its promising applications in high-power and high-frequency electronics, such as white light emitting diodes (LEDs),[6-8] high electron mobility transistors (HEMTs),[9-11] and high power transmissions.[12-13] Among its many useful properties, thermal conductivity is critical for the stable performance and safe operation of SiC devices at high temperatures, high frequency, and high voltages. While the thermal conductivity of SiC has been reported previously,[14-21] conflicting data still exists among different works. For example, it remains controversial whether the thermal conductivity of 4H phase of SiC is higher than that of the 6H phase.[15, 19-20] More importantly, while the anisotropy in the thermal conductivity of both 4H- and 6H- SiC is expected due to their hexagonal Bravais lattice structures (as shown in Figure 1a), it has usually been ignored in previous experimental studies[14, 17-18] due to the challenges in accurate measurements. So far, there is only one experimental work[16] reporting the anisotropic thermal conductivity of 6H-SiC measured using photothermal radiometry, which shows that the cross-plane thermal conductivity $k_z$ (perpendicular to the hexagonal planes) of 6H-SiC is 30% lower than its in-plane thermal conductivity $k_r$ (parallel to the hexagonal planes). The anisotropic thermal conductivity of 4H-SiC has not been systematically studied experimentally.

In this paper, we use the time-domain thermoreflectance (TDTR)[22] to simultaneously determine both the $k_r$ and $k_z$ of three SiC single crystals provided by II-VI Inc.®: unintentionally doped (UID) semi-insulating (SI) 4H-SiC, n-type 4H-SiC (N-doped $1\times10^{19}$ cm$^{-3}$), and SI 6H-SiC (V-doped $1\times10^{17}$ cm$^{-3}$), over a temperature range from 250 K to 450 K. Anisotropy is observed in thermal conductivity of all the SiC samples, with $k_z$ ~40% lower than $k_r$. For both $k_z$ and $k_r$, the measured thermal conductivity has the following decreasing order: SI 4H-SiC > n-type 4H-SiC > 6H, which agrees well with the recent first principles predictions by Protik *et al*.[19]

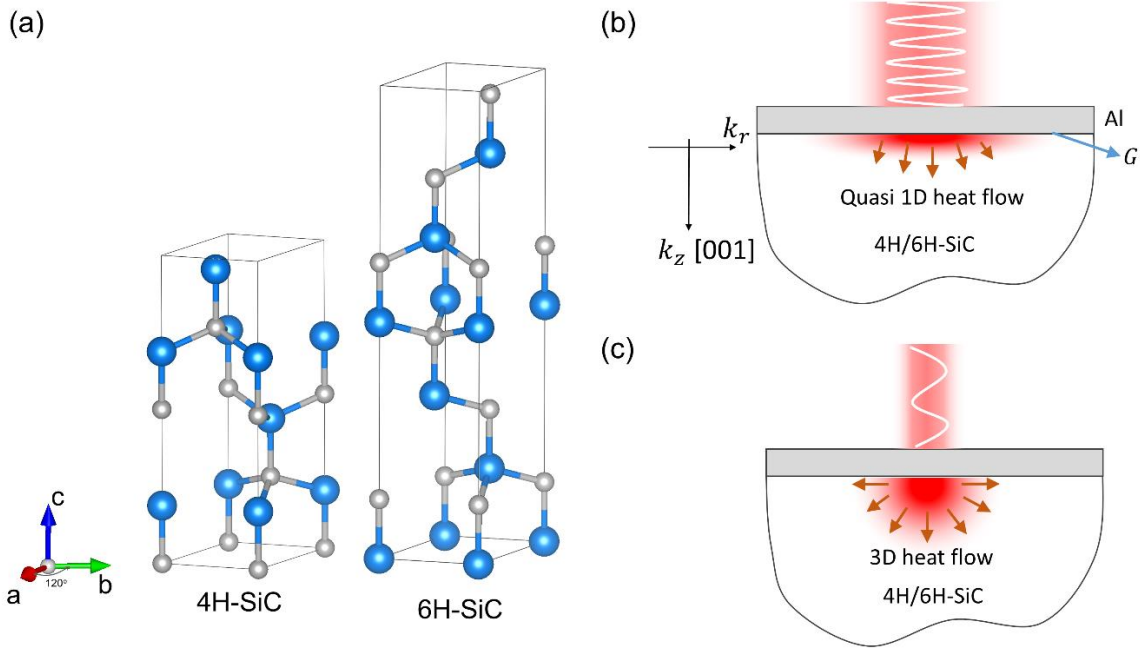

Figure 1. (a) Atomic structure of 4H-SiC and 6H-SiC. (b) Schematic for measuring $k_z$ using a large spot size and a high modulation frequency of TDTR measurements. (c) Schematic for measuring $k_r$ using a small spot size and a low modulation frequency of TDTR measurements.

We measure the anisotropic thermal conductivity $k_r$ and $k_z$ using TDTR by varying the laser spot size and the modulation frequency,[23] as shown in Figure 1b-c, where a ~110 nm Al transducer was deposited on all SiC samples. In TDTR measurement, the surface heating by the laser is transient and non-uniform with a Gaussian profile. Although the heat conduction with such transient Gaussian heating as boundary condition can be fully solved in the cylindrical coordinates,[24] it is still necessary to understand how to appropriately choose the experimental parameters for separate measurement of the anisotropic thermal conductivity $k_r$ and $k_z$. There are two important length scales in the TDTR experiments that determine the heat flow direction in the SiC substrate and hence the different sensitivities to $k_r$ and $k_z$. The first length scale is the size of the Gaussian laser spot, defined as the root-mean-square average of the $1/e^2$ radii of the pump ($w_0$) and the probe ($w_1$) as: $w = \sqrt{(w_0^2 + w_1^2)/2}$. The other important length scale is the thermal penetration length:

$$d_{P,\alpha} = \sqrt{k_\alpha/C\pi f_0} \tag{1}$$

where the subscript $\alpha (= r, z)$ denotes the direction in cylindrical coordinates, $k$ is the thermal conductivity, and $C$ is the volumetric heat capacity. Since TDTR measures the surface temperature rise within the RMS radius of the laser spot, whether the TDTR signal is sensitive to $k_r$ depends on how large the laser spot radius $w$ is compared to the in-plane thermal diffusion length $d_{P,r}$. If the spot radius $w$ is much larger than the in-plane penetration length $d_{P,r}$, the in-plane temperature gradient is negligible and the heat flow can be regarded as one-dimensional along the cross-plane direction. Based on our previous work, the criterion for satisfying the quasi one-dimensional thermal transport along the cross-plane direction is: [23]

$$w \geq 5 d_{P,r} \tag{2}$$

By satisfying Eq. (2), the TDTR signal is only sensitive to the parameters associated with cross-plane heat transfer, namely, the cross-plane thermal conductivity $k_z$ and the interface conductance $G$ between the transducer and the SiC substrate. After determining $k_z$ and $G$, $k_r$ can then be measured reliably if the laser spot radius is chosen to satisfy the condition that the in-plane penetration length $d_{P,r}$ is at least half of the laser spot radius:[23]

$$d_{P,r} \geq \frac{1}{2} w \tag{3}$$

In principle, both the modulation frequency $f_0$ and the laser spot radius $w$ can be varied to separately measure $k_r$ and $k_z$.[25] For example, $k_z$ could be first measured using a large spot radius $w$ at a high modulation frequency $f_0$ (as shown in Figure 1b), then $k_r$ can be measured using a small spot radius $w$ at a low modulation frequency (as shown in Figure 1c). However, we need to be cautious that the measured $k_z$ could depend on $f_0$ when different phonon modes are out of thermal equilibrium [26-31] and $k_r$ could be underestimated if $w_0$ is smaller than the mean free paths of heat carrying phonons.[29, 32-34]

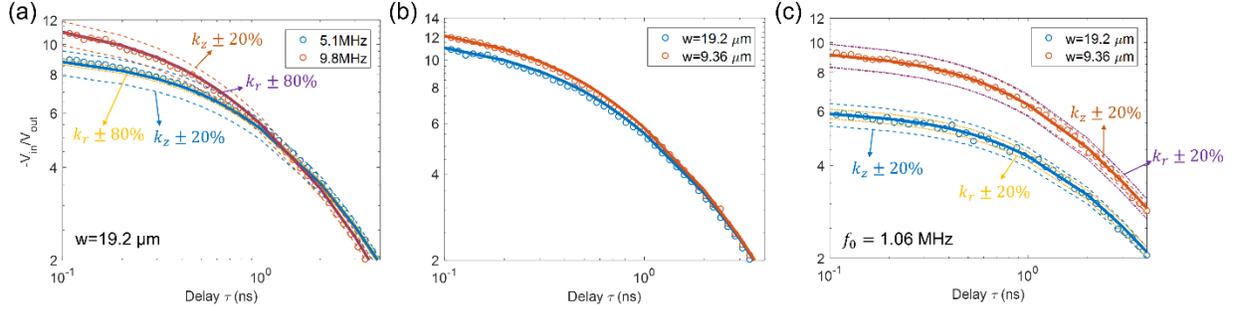

Figure 2. (a) Measurement of cross-plane thermal conductivity $k_z$ of SI 6H-SiC sample at room temperature with $f_0 = 5.1$ MHz and 9.8 MHz where the TDTR signal has negligible sensitivity to $k_r$. The best parameters are $k_z = 273$ W/mK and $G = 128$ MW/m²K. (b) The best-fit $(k_z, G)$ obtained using $w = 19.2$ μm can also fit the TDTR signal using a smaller spot radius $w = 9.4$ μm, indicating $k_z$ and $G$ are independent of laser spot size. (c) Measurement of in-plane thermal conductivity $k_r$ at low modulation frequency at $f_0 = 1.06$ MHz. Using both $w = 9.4$ μm and $w = 19.2$ μm, the obtained in-plane thermal conductivity is the same $k_r = 393$ W/mK.

We perform the following measurements to make sure that the measured thermal conductivity of SiC samples is intrinsic and is not affected by the choices of operation parameters such as the laser spot radius or the modulation frequency. First, we check whether $k_z$ depends on $f_0$ by using a large spot radius $w = 19.2$ μm. Using the in-plane thermal conductivity $k_r$ from first principles[19] calculation and heat capacity $C$ taken from ref [35], we estimate from Eq. (2) that the modulation frequency $f_0$ should be higher than 4.6 MHz for the independent measurement of $k_z$. We thus conduct the measurements for $k_z$ at two different modulation frequencies of $f_0 = 5.1$ MHz and 9.8 MHz, with the obtained signals shown in Figure 2a. It is clear that the obtained signal can be regarded as independent of $k_r$. To fit the experimental signal, nonlinear least-squares regression is used in this work. The cost function for the regression is defined as:

$$W(\boldsymbol{U}) = \sum_j \sum_i [R_{Exp}(\tau_i, f_{0j}) - F(\tau_i, f_{0j}, \boldsymbol{U}, \boldsymbol{P})]^2 \qquad (4)$$

where $R_{Exp}(\tau_i, f_{0j})$ is the ratio $-V_{in}/V_{out}$ between the in-phase signal $V_{in}$ and the out-of-phase signal $V_{out}$ measured at delay time $\tau_i$ and modulation frequency $f_{0j}$, and the function $F$ denotes the full solution of heat conduction equation with periodic Gaussian heating profile as the boundary condition, which is used to predict $-V_{in}/V_{out}$. We note that the ratio $-V_{in}/V_{out}$ is essentially equivalent to the phase $\phi$ of the signal since $\tan\phi = V_{out}/V_{in}$. The vector $\boldsymbol{P}$ is the set of control parameters including the thickness, heat capacity and thermal conductivity of the Al transducer, as well as the laser spot radius. The vector $\boldsymbol{U} = [k_r, k_z, G]^T$ is a set of unknown parameters that need to be determined during the nonlinear regression, where $G$ is the interface conductance between the SiC and the Al transducer. As discussed above, since $k_r$ has negligible effect on the predicted ratio $-V_{in}/V_{out}$ when Eq. (2) is satisfied, $U$ can be reduced to $\boldsymbol{U} = [k_z, G]$ and $k_r$ is set equal to $k_z$ during the nonlinear regression. Using the simplex algorithm,[36] the unknown parameters $\boldsymbol{U}$ are adjusted iteratively, until the change of an element in $\boldsymbol{U}$ and the reduction of $W$ between the succeeding steps are both smaller than 0.1%. Using the nonlinear regression, we found that the signals obtained at 5.1 MHz and 9.8 MHz can be fitted with the same value of $k_z$ and $G$, indicating that the cross-plane thermal transport in SiC is not affected by the modulation frequency. We then performed the measurement for $k_z$ at the same modulation frequency of $f_0 = 9.8$ MHz using different spot radii of $w = 19.2$ μm and 9.36 μm, which yield the same $k_z$, indicating that the cross-plane thermal conductivity $k_z$ is not affected by the spot radius $w$ either, as shown in Figure 2b.

After making sure that $k_z$ and $G$ are not affected by $f_0$ or $w$, we proceed to measure the in-plane thermal conductivity $k_r$. Using Eq. (3), we determine that the modulation frequency should be $f_0 \leq 2.9$ MHz to ensure that the measured signal is sensitive to $k_r$ when we use the spot radius $w = 9.4$ μm. We therefore select $f_0 = 1.06$ MHz to measure $k_r$, as shown in Figure 2c. Similar to the cross-plane thermal conductivity measurement, we use the same nonlinear regression algorithm, but the unknown parameter $\boldsymbol{U} = k_r$ in this case, and $k_z$ and $G$ are regarded as known parameters by grouping them into $\boldsymbol{P}$. We then repeated the measurement described above but with a larger spot radius $w = 19.2$ μm to make sure the $w$ is large enough so that $k_r$ reaches the diffusive limit. We found the $k_r$ measured using $w = 9.4$ μm can successfully fit the TDTR signal obtained with $w = 19.2$ μm, indicating the measured $k_r$ is already converged with respect to $w$. Based on the first-principles calculation,[19] the mean free paths of heat carrying phonons are estimated to be in the range 1~12 μm at room temperature. The smallest diameter of we use in our measurement $2w = 18.8$ μm is still larger than the longest phonon mean free paths, and the ballistic effect induced by the limited laser spot size should be negligible. Through the above measurements, we are confident that both $k_r$ and $k_z$ of 6H-SiC are intrinsic values, free of any extra error induced by ballistic transport or non-equilibrium transport. We performed similar measurements on SI and n-type 4H-SiC samples, and we obtained robust $k_r$ and $k_z$ independent of the modulation frequency and the laser spot radius. In the rest of the paper, we therefore measure $k_z$ at $f_0 = 9.8$ MHz and $k_r$ at $f_0 = 1.06$ MHz for all the SiC samples using the same spot radius $w = 9.4$ μm.

The discussion above suggests that the unknown parameters $\boldsymbol{U} = [k_r, k_z, G]^\mathrm{T}$ can be simultaneously determined through fitting the signal obtained at $f_0 = 1.06$ MHz and 9.8 MHz at the same spot radius $w = 9.4$ μm. Since we are measuring multiple parameters at multiple modulation frequencies, an error propagation formula based on the least-squares regression is necessary. We extended the error propagation formula by Yang *et al.*[37] in our previous work for the case when multiple modulation frequencies are used.[31] The error propagation formula is written as:

$$var[\boldsymbol{U}] = \boldsymbol{\Sigma}_U^{-1}\left[\sum_j \boldsymbol{J}_U^T(f_{0j})var[\boldsymbol{R}_{Exp}(f_{0j})]\boldsymbol{J}_U(f_{0j})\right]\boldsymbol{\Sigma}_U^{-1} + \boldsymbol{\Sigma}_U^{-1}\boldsymbol{\Sigma}_{UP}var[\boldsymbol{P}]\boldsymbol{\Sigma}_{UP}^T\boldsymbol{\Sigma}_U^{-1} \quad (5)$$

where $var[\cdot]$ denotes the covariance matrix, $\boldsymbol{R}_{Exp}(f_{0j}) = \left[-\frac{V_{in}}{V_{out}}(\tau_1, f_{0j}), \ldots, -\frac{V_{in}}{V_{out}}(\tau_i, f_{0j}), \ldots\right]^T$ is the vector containing the TDTR ratio between in-phase signal $V_{in}$ and out-of-phase signal $V_{out}$ measured at a sequence of delay time $[\tau_1, \tau_2, \ldots, \tau_i, \ldots]^T$ and modulation frequency $f_{0j}$, and, $\boldsymbol{P}$ is the vector containing input parameters including the rms laser spot radius $w$, thickness $d_{Al}$, heat capacity $C_{Al}$ and thermal conductivity $k_{Al}$ of the transducer, and the heat capacity $C$ of SiC. The $\boldsymbol{\Sigma}$ matrices in Eq. (5) are written as:

$$\boldsymbol{\Sigma}_U = \sum_j \boldsymbol{J}_U^T(f_{0j})\boldsymbol{J}_U(f_{0j}), \quad \boldsymbol{\Sigma}_{UP} = \sum_j \boldsymbol{J}_U^T(f_{0j})\boldsymbol{J}_P(f_{0j}) \quad (6)$$

where $\boldsymbol{J}_U(f_{0j}) = \left.\frac{\partial(F(\tau_1),F(\tau_2),\ldots,F(\tau_i),\ldots)}{\partial(k_r,k_z,G)}\right|_{f_{0j}}$ and $\boldsymbol{J}_P(f_{0j}) = \left.\frac{\partial(F(\tau_1),F(\tau_2),\ldots,F(\tau_i),\ldots)}{\partial(w,d_{Al},C_{Al},k_{Al},C)}\right|_{f_{0j}}$ are the Jacobi matrices of the thermal model $F$ with respect to $\boldsymbol{U}$ and $\boldsymbol{P}$ at frequency $f_{0j}$, respectively. The diagonal elements in the $var[\boldsymbol{U} \text{ or } \boldsymbol{P}]$ are essentially the variance $\sigma^2$ of the parameters, and we use $2\sigma$ as the uncertainty reported in this paper. The uncertainties ($2\sigma$) of the control parameters $\boldsymbol{P}$ are estimated as follows: 10% for the thermal conductivity of Al, 5% for the heat capacity of Al and the substrate, 5% for the Al thickness, and 4% for the laser spot size.[23, 31] We summarize the calculated $4 \cdot var[\boldsymbol{U}]$ for the SI 4H-SiC, n-type 4H-SiC and SI 6H-SiC in Table 1, so that the uncertainties $2\sigma$ can be directly calculated as square root of the diagonal elements.

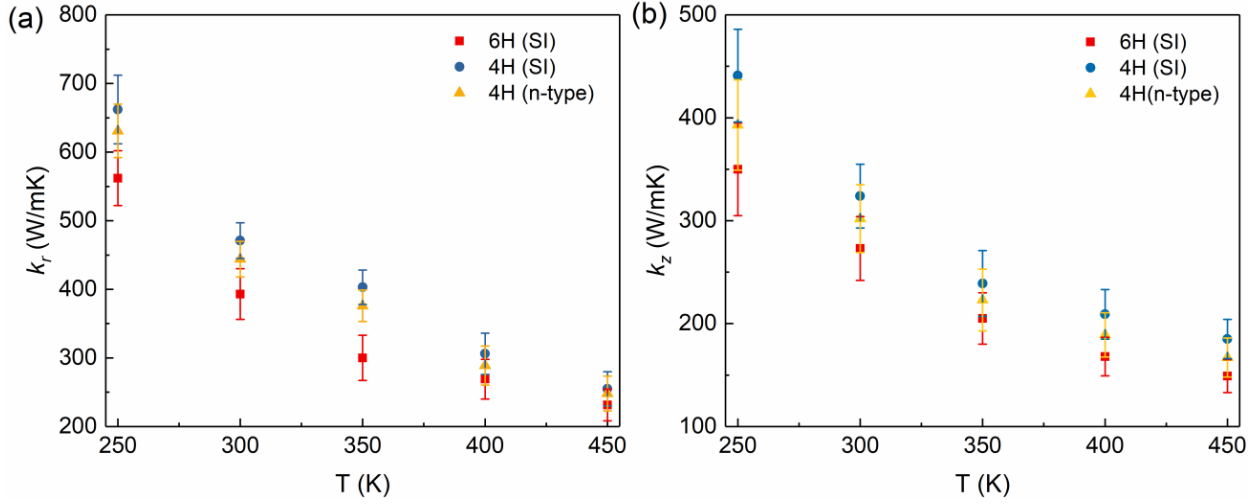

Figure 3. Temperature-dependent thermal conductivity of SI 4H SiC, n-type 4H SiC ,and SI 6H SiC in (a) the in-plane direction and (b) the cross-plane direction.

Figure 3 summarizes the temperature-dependent $k_r$ and $k_z$ for the SI 4H-SiC, n-type SiC, and SI 6H-SiC from 250 K to 450 K. Anisotropy is clearly observed in the measured thermal conductivity for all three SiC samples from 250 K to 450 K, with $k_z$ about 40% lower than $k_r$. The SI and n-type 4H-SiC have higher $k_r$ and $k_z$ than those of SI 6H-SiC sample, which agree well with the first principles predictions that the thermal conductivity of $n$H-SiC ($n = 2, 4, 6$) decreases with increasing $n$.[19] The SI 4H-SiC has the highest thermal conductivity among the three SiC crystals, with 7% higher thermal conductivity than the n-type 4H-SiC due to the phonon-impurity scattering in n-type 4H SiC. Figure 4 shows the interface conductance between the Al transducer and the three SiC samples. Because of the high Debye temperature of SiC (4H 1300 K and 6H 1200 K),[38] the interface conductance of three samples increases as the temperature rises from 250 K to 450 K. The 6H SiC has higher interface conductance with Al, because its Debye temperature is better matched with Al (433 K).[39]

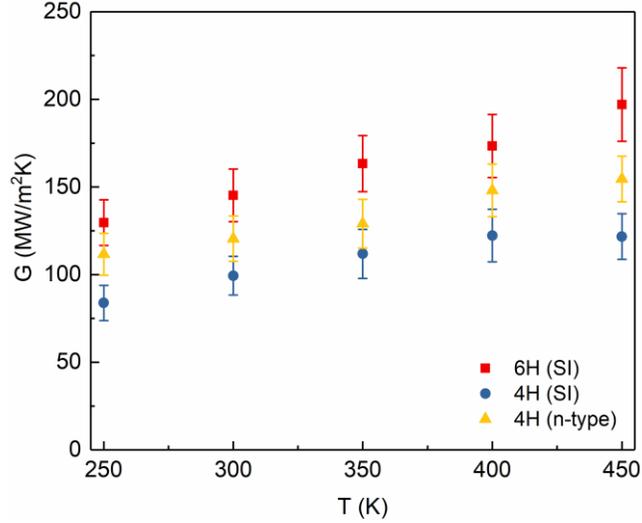

Figure 4. Interface conductance between Al transducer and the SiC samples.

Figure 5 compares our measured $k_r$ and $k_z$ for 4H-SiC with relevant experimental measurements and the first-principles calculation results in literature. In the in-plane direction, our results agree well with the first principles calculation,[19] but much higher than the measurement by Morelli et al.[20] The much lower thermal conductivity by Morelli *et al.* [20] is due to the defects in their 4H-SiC samples, as suggested by the authors. In the cross-plane direction, the measured $k_z$ for SI 4H-SiC is slightly smaller than the first-principles calculations but higher than the laser flash analysis (LFA) measurements by Wei *et al.*[17] above 350 K. It is confusing that the temperature-dependent thermal conductivity measured in their work has the $1/T^2$ temperature dependence, largely deviating from the $1/T$ law.[40] In Figure 6, we compare the anisotropic thermal conductivity of SI 6H-SiC with both first principles calculations and the measurements by others. For both the $k_r$ and $k_z$, our TDTR measurements agree well with the first-principles calculation[19] and measurement by others. [16, 19, 21]

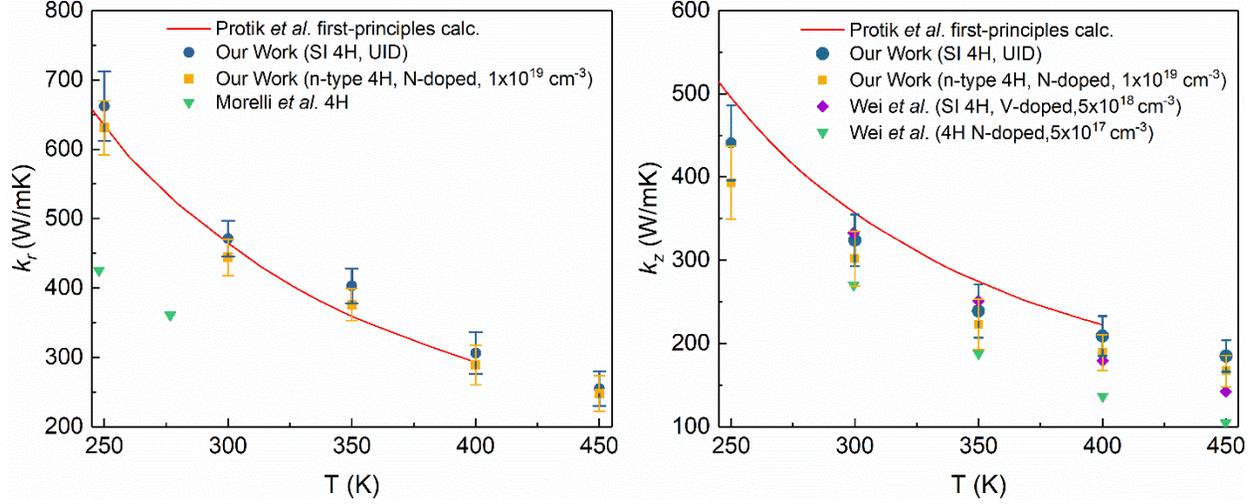

Figure 5. (a) In-plane thermal conductivity for SI and n-type 4H-SiC compared with the first-principles calculation by Protik et al.,[19] and the steady-state measurement by Morelli et al.[20] (b) The cross-plane thermal conductivity for SI and n-type 4H-SiC compared with the calculation by Protik et al.[19] and the laser flash analysis measurement by Wei et al.[17]

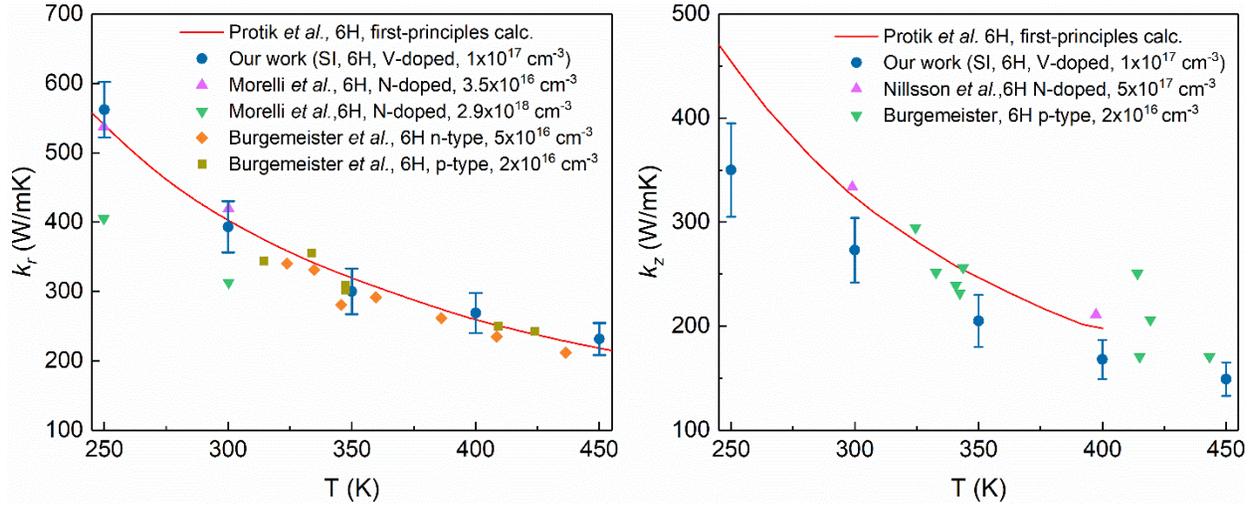

Figure 6 (a) In-plane thermal conductivity for SI 6H-SiC compared with the first-principles calculation by Protik et al.,[19] the steady-state measurement by Morelli et al.[18] and the radiation thermometry by Burgemeister et al.[16] (b) The cross-plane thermal conductivity for SI 6H-SiC compared with the calculation by Burgemeister et al.[16], Protik et al.,[19] and Nilsson et al.[21]

In summary, we have measured both the in-plane and the cross-plane thermal conductivity of SI 4H-SiC, n-type 4H-SiC, and SI 6H-SiC using TDTR by varying both the laser spot radius and the modulation frequency of TDTR measurements. We developed a measurement protocol to make sure that the measured thermal conductivities are intrinsic values, independent of the choices of operational parameters such as the laser spot radius and the modulation frequency. Our measurement results confirmed the first-principles prediction that thermal conductivity is anisotropic in the hexagonal SiC crystals, and that 4H-SiC has higher thermal conductivity than 6H-SiC. This work provides an important benchmark for understanding thermal transport in WBG semiconductors.

**Acknowledgements.** This work was supported by the NSF (Grant No. 1512776). RY acknowledges Rajan Rengarajan at II-VI Inc. and John Blevins at AFRL for providing the SiC samples and helpful discussions.

**Table 1**. The covariance matrices $4 \cdot var[k_r, k_z, G]$ for the SiC samples at room temperature. The diagonal of the matrices shown in the table is the uncertainty level $2\sigma$. The units shown for the covariance matrices are $k_r$ (W/mK), $k_z$ (W/mK) and $G$ (MW/m²K).

| SI 6H- SiC | $k_r$ | $k_z$ | $G$ |
|---|---|---|---|
| $k_r$ | 1101 | -73.7 | 21.3 |
| $k_z$ | -73.7 | 1267 | 302.2 |
| $G$ | 21.3 | 302.2 | 94.6 |
| Best-fit | 393 | 273 | 128 |
| Uncertainty | 8.4% | 13.0% | 7.6% |
| SI 4H-SiC | $k_r$ | $k_z$ | $G$ |
| $k_r$ | 1588.1 | -372.2 | 26.8 |
| $k_z$ | -372.2 | 2421.0 | 275.8 |
| $G$ | 26.8 | 275.8 | 53.4 |
| Best-fit | 471 | 324 | 101 |
| Uncertainty | 8.5% | 15.2% | 7.2% |
| n-type 4H-SiC | $k_r$ | $k_z$ | $G$ |
| $k_r$ | 1315.5 | -206.4 | 25.1 |
| $k_z$ | -206.4 | 1642.2 | 284.3 |
| $G$ | 25.1 | 284.3 | 76.3 |
| Best-fit | 444 | 302 | 121 |
| Uncertainty | 8.7% | 13.4% | 7.3% |